\documentclass[a4paper,11pt]{article}
\usepackage{jinstpub} 
\usepackage{lineno}

\title{Recent GasPM advances: photon-feedback mitigation and LaB$_{6}$ photocathode studies}

\author[a]{S. Garnero}
\author[bce]{, K. Inami}
\author[bcde]{, K. Matsuoka}
\author[a]{, R. Okubo}
\author[b]{, K. Ueda}

\affiliation[a]{INFN Sezione di Trieste, Italy}
\affiliation[b]{Nagoya University, Nagoya, Japan}
\affiliation[c]{High Energy Accelerator Research Organization (KEK), Tsukuba, Japan}
\affiliation[d]{The Graduate University for Advanced Studies (SOKENDAI), Hayama, Japan}
\affiliation[e]{Kobayashi-Maskawa Institute for the Origin of Particles and the Universe (KMI), Nagoya, Japan}

\emailAdd{simone.garnero@ts.infn.it}

\abstract{
We report recent developments and tests with beams and cosmic rays of the gaseous photomultiplier (GasPM). The GasPM is a photosensor that combines a photocathode with the avalanche-multiplication mechanism of a resistive-plate chamber, offering excellent time resolution and cost-effective scalability. In addition, the GasPM provides precise and efficient Cherenkov-based charged-particle identification if combined with a radiator. Our primary use case aims at an upgrade of the Belle II detector to suppress beam-induced background photons, preferably detected off-collision time, that degrade the performance of the electromagnetic calorimeter. In 2022 we achieved a promising single-photon time-resolution of 25 ps at 3.3 × 10$^6$ gain, using a picosecond-pulse laser and a LaB$_6$ photocathode. However, a 2023 beam test with electrons impinging on a MgF$_2$ window attached to a CsI photocathode showed a worsening to 70 ps.
This work aims at addressing the principal causes of the time-resolution degradation. We primarily target ultraviolet-photon emission during excitation and de-excitation of the gas molecules, which leads to a secondary signal that overlaps the primary signal, spoiling time resolution (photon feedback). We design and execute an improved beam test. Along with several GasPM configuration changes, we introduce a new 10 GSPS frequency digitizer to better discriminate primary from secondary signals thus enabling the study of photon feedback. We also conduct a cosmic-ray test using a LaB$_6$ photocathode, which is known to have higher than CsI's resistance to ions drifting backwards onto the photocathode and to air exposure, to probe quantum efficiency in view of an upcoming beam test.
}

\begin{document}
\maketitle
\flushbottom

\section{Introduction}
\label{sec:intro}

Belle II is an hermetic magnetic spectrometer surrounded by particle-identification detectors and a calorimeter~\cite{TDC}. It studies the decays of billions of $B$, $D$, and $\tau$ particles from $e^+e^-$ collisions at the $\Upsilon$(4S) energy, produced by the SuperKEKB collider at KEK~\cite{Akai_2018}.
The CsI crystal calorimeter is central in the Belle II physics program, as its hermeticity and precise detection of photons and electrons enable unique key measurements. However, its proximity to the beams, technology, and large acceptance make it sensitive to beam-background photons. These are energy deposits not associated with the primary physics $e^+e^-$ interaction. They originate from various beam-induced processes, including radiative Bhabha scattering and single-beam interactions with residual gas in the vacuum~\cite{background}. The latter are photons associated with the harsh and tightly constrained collision environment required by the SuperKEKB nano-beam scheme~\cite{nanobeam}. Their energies are typically 1--2 MeV but can reach sufficient values to be reconstructed as calorimeter clusters. Because the collision time is precisely known from the SuperKEKB radio-frequency, a promising strategy to cope with beam-backgrounds is to exploit detection-time to recognise photons not originated from the collision, which are preferentially out-of-time.
We developed the gaseous photomultiplier (GasPM)~\cite{gaspm}, a photodetector, based on a resistive-plate chamber (RPC) coupled with a photocathode, that is expected to offer $\mathcal{O}(10)$~ps time resolution over large coverage at low cost.
In addition, paired with a radiator, the GasPM operates as a Cherenkov detector providing an affordable alternative to devices like micro-channel-plate (MCP)-PMTs~\cite{top_mcp_pmt} for large area applications.

\section{Design}
\label{sec:design_first_proto}
\begin{figure}[htbp]
    \centering
    \hspace*{-0.3cm} 
    \includegraphics[width=1.05\linewidth]{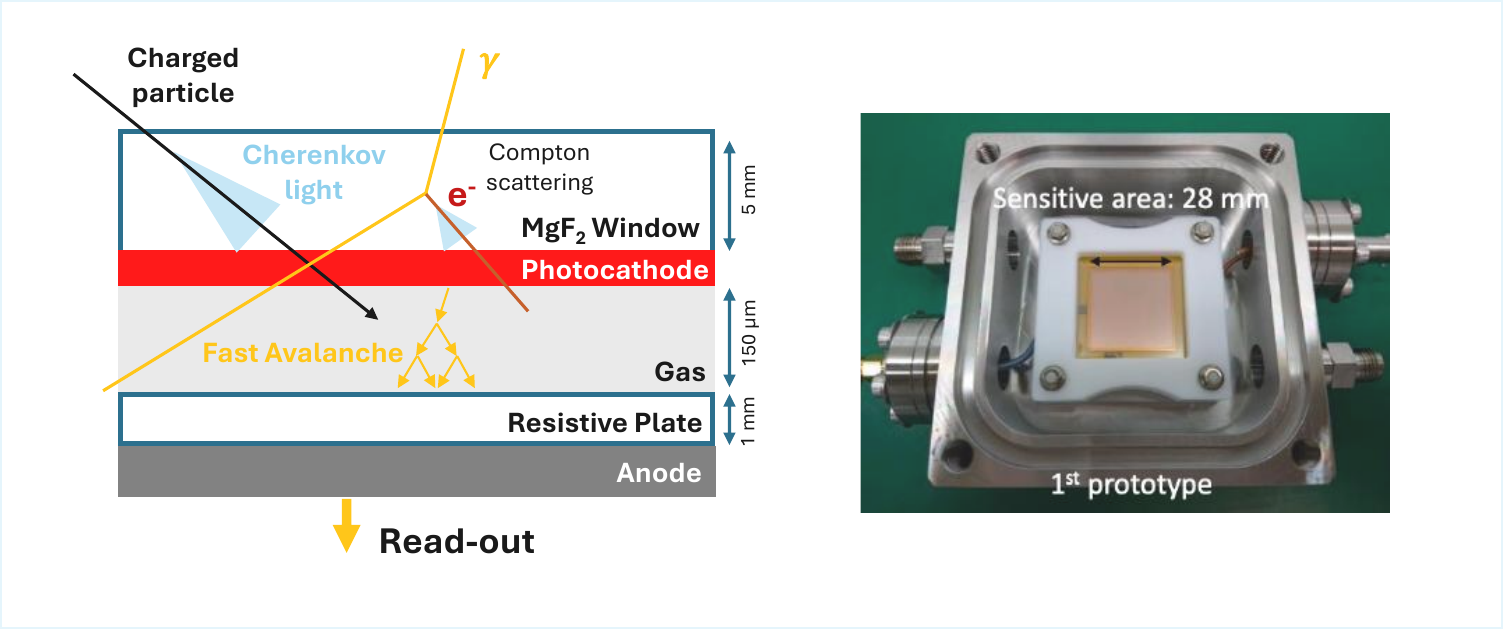}
    \caption{Design of the GasPM prototype (left) and a photo of the prototype (right).}
    \label{fig:gaspm_1st}
\end{figure}

The GasPM architecture is shown in Fig.~\ref{fig:gaspm_1st}. 
Photons pass through a transparent window and hit the photocathode. In our specific use case, few-MeV photons undergo Compton scattering in the window and the resulting electron emits Cherenkov photons that then hit the photocathode. Charged particles simply emit Cherenkov photons in the window.
Photoelectrons emitted from the photocathode get accelerated by the adjacent electric field and generate a Townsend avalanche in the narrow gas gap, filled with a mixture of $90\%$ $\rm{C_2 H_2F_4}$ and $10\%$ $\rm{SF}_6$ gases. The charge generated in the avalanche drifts, inducing a signal at the copper electrode. The photocathode is mounted on a transparent window and a soda-glass resistive plate prevents discharges. The window and photocathode material, applied voltage, gap thickness, gas admixture, and restive plate can be optimized to each specific application.

\section{Previous results}
\begin{figure}[htbp]
    \centering
    \begin{minipage}{0.48\textwidth}
        \centering
        \includegraphics[width=\textwidth]{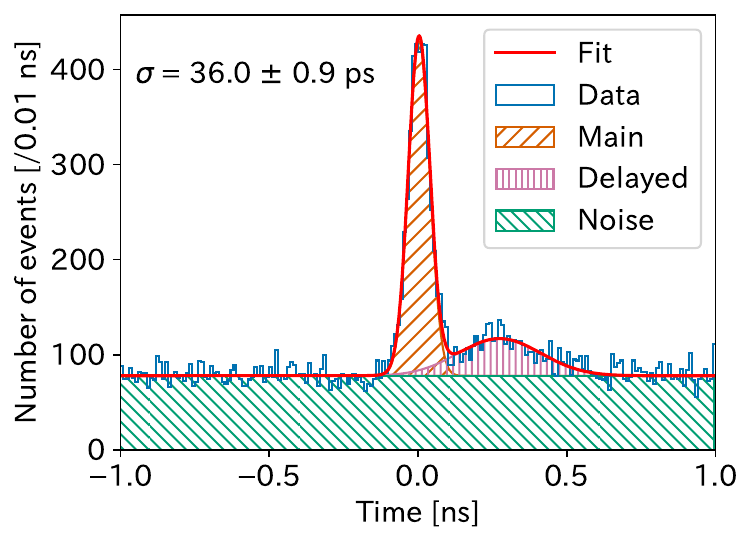}
    \end{minipage}
    \hfill
    \begin{minipage}{0.48\textwidth}
        \centering
        \includegraphics[width=0.9\textwidth]{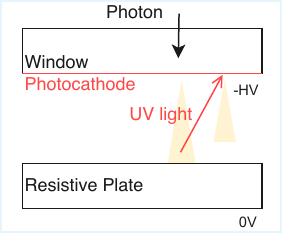}
    \end{minipage}
    \caption{Single-photon time resolution from laser test~(left)~\cite{gaspm} and a schematic of the photon-feedback mechanism~(right).}
    \label{fig:resol}
\end{figure}

In 2022 a first test achieved good time resolution. Using a 375~nm picosecond-pulse laser; a quartz window; an LaB$_6$ photocathode, which has low quantum efficiency but stable performance in gases; a 170~$\mu$m gas gap for 176 kV/cm electric field; and a 1.1~mm thick TEMPAX float glass, the GasPM achieved a single-photon time resolution of $\sigma = \rm{25.0\pm1.1~ps}$ (Fig.~\ref{fig:resol}~(left))~\cite{gaspm}. A broad "shoulder" signal delayed by 0.1--0.5 ns was observed. 
In 2023 we developed a prototype to demonstrate the Cherenkov application. We used a CsI photocathode, known to have a sufficient quantum efficiency for UV light and a good tolerance to the relevant gas admixture. We also used a 2.4~mm $\rm{MgF_2}$ window for better UV light transmission. The thickness of the gas gap was $200~\rm{\mu m}$ with an electric field of only 140~kV/cm. The resulting reduced electrons drift velocity was therefore expected to worsen the time resolution.
We tested it using a 3~GeV electron beam. The majority of signals presented overlapping pulses within a time window of roughly $600$~ns, which made the timing determination difficult. The resulting time resolution was $\sigma = 73.0\pm2.4~\rm{ps}$.
These measurements exposed two main GasPM limitations. Photon-feedback, sketched in Fig.~\ref{fig:resol}~(right), are photons emitted during gas excitation and de-excitation that hit the photocathode, triggering secondary avalanches in the gas gap. Photon feedback generates delayed signals that overlap with the primary ones, degrading time resolution. Ion feedback are avalanche ions in the gas gap that drift backwards and hit the photocathode, damaging it and thus degrading the efficiency over time.

\begin{figure}[htbp]
    \centering
    \includegraphics[width=\linewidth]{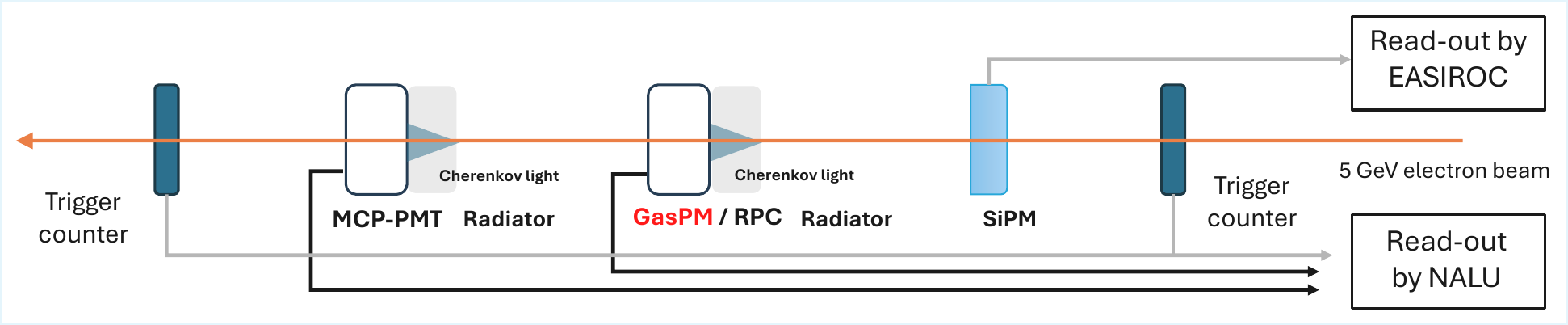}
    \caption{Schematic of the beam-test setup at KEK PF-AR test beam-line.}
    \label{fig:beam_test}
\end{figure}

\section{Photon feedback study in a beam test}
In this work we report on an improved beam test with modifications aimed at enhancing time resolution by means of understanding and possibly addressing photon feedback. We increase the electric field to 187~kV/cm by reducing the gas gap to 150~$\mu$m, and we increase the MgF$_2$ window thickness to 5~mm to increase the photon yield.
To this end we improve the read-out rate from 5~GSPS to 10~GSPS using a new digitiser, the Nalu DSA-C10-8+~\cite{NALU}. The goal is to better discriminate primary signals from secondary pulses originating from photon feedback. We also change the resistive plate material to soda glass.
We use a 5~GeV electron beam at the KEK PF-AR test beam-line. Figure~\ref{fig:beam_test} shows the experimental configuration. Together with the GasPM, we use an MCP-PMT with a quartz radiator as a time reference, a SiPM array with an acrylic radiator to veto $\delta$-ray–induced multi-electron events, and two plastic scintillators read out by PMTs as trigger counters along the beam line. 

\begin{figure}[htbp]
    \centering
    \begin{minipage}{0.48\textwidth}
        \centering
        \includegraphics[width=\textwidth]{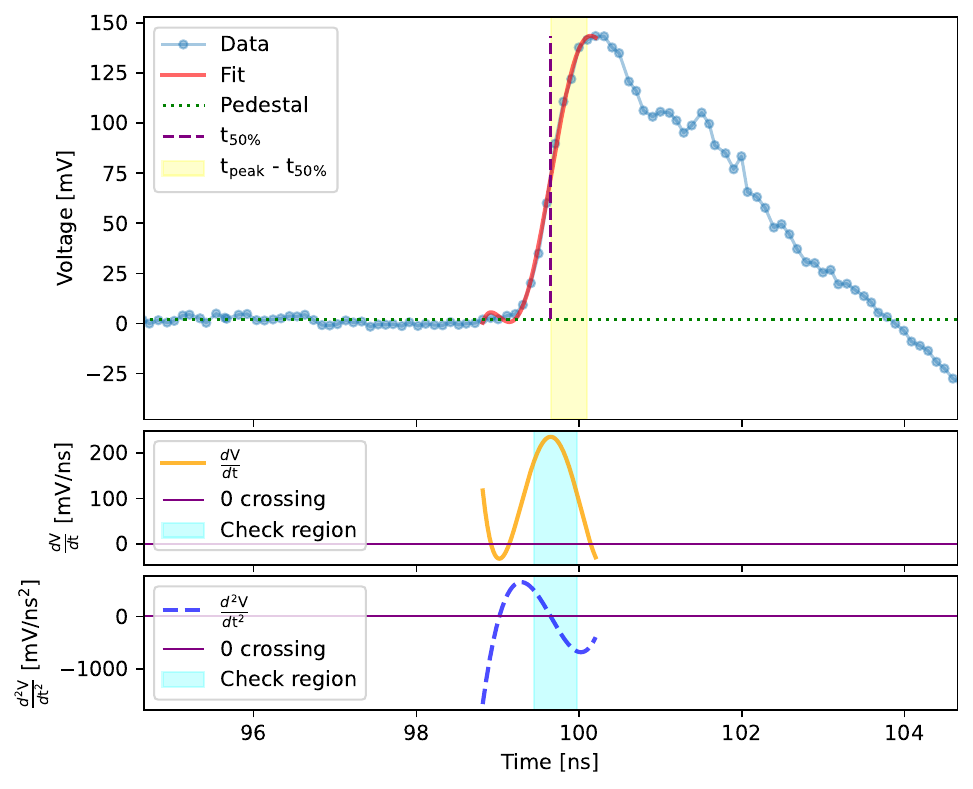}
    \end{minipage}
    \hfill
    \begin{minipage}{0.48\textwidth}
        \centering
        \includegraphics[width=\textwidth]{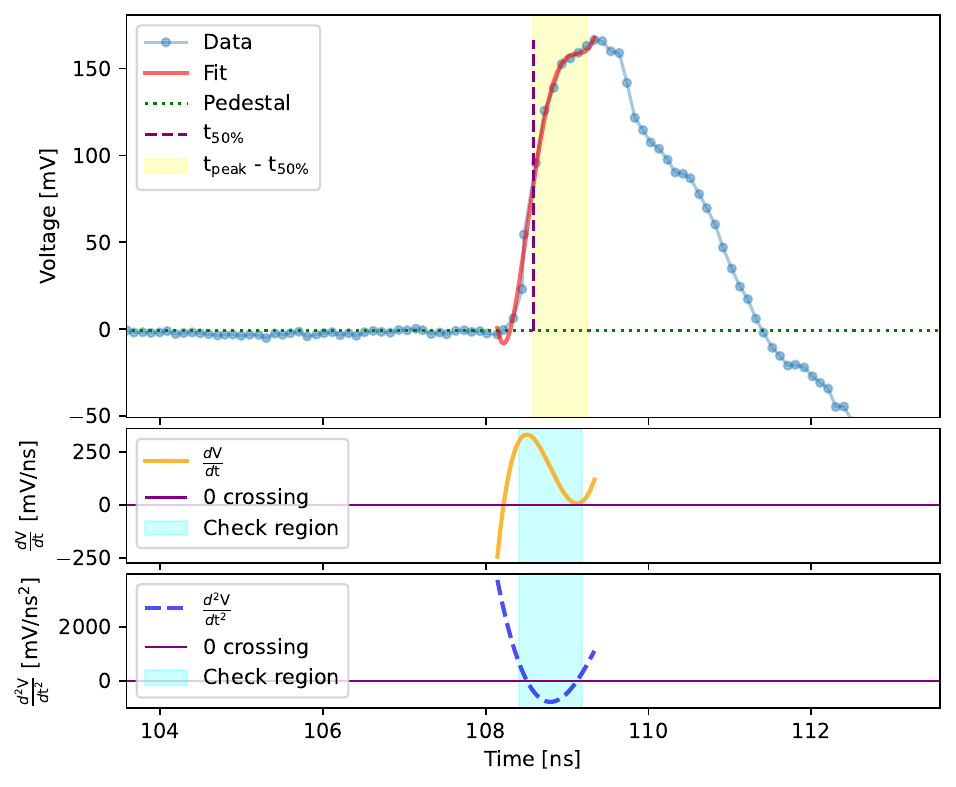}
    \end{minipage}
    \caption{Waveforms from two beam-test events with polynomial fit projections overlaid and the relevant derivatives in the bottom panels. One is tagged as single avalanche (left) and the other as photon feedback (right).}
    \label{fig:PF}
\end{figure}

\begin{figure}[htbp]
    \centering
    \includegraphics[width=0.6\linewidth]{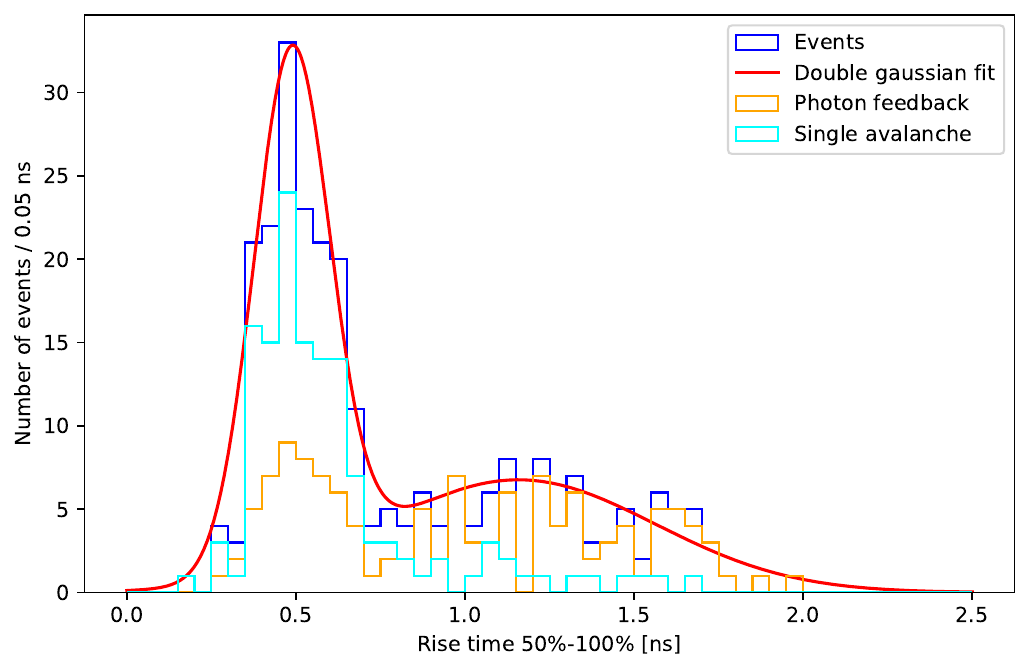}
    \caption{Distribution of the pulse rise time from 50\% to the peak after applying the algorithm. Events in cyan and orange are respectively tagged as single avalanche and photon-feedback events.}
    \label{fig:PF_selection}
\end{figure}

Classification of beam-test events is not trivial because many processes occur and the direct correlation between pulse height, charge, and photon feedback, typical of the laser test, is diluted. Moreover, the thinner gas gap increases overlap between secondary and primary signals.
We therefore focus on the rising edge of the signals and develop an algorithm to distinguish photon-feedback from single-avalanche events.
We define the rising edge of the waveform as the data from the pedestal to the pulse peak, and fit it with a 8th-order polynomial. We then select events exhibiting at least two zero-crossings in the second derivative of the fit within the rising edge, excluding a 0.2 ns fixed margin from
the edge of the pedestal level, and another 0.2 ns fixed margin before the peak, to avoid boundary artifacts. The presence of multiple zero-crossings indicates different curvatures, inconsistent with a single-component rise. Figure~\ref{fig:PF} shows the waveforms of two events, one tagged as single avalanche (left) and the other as photon feedback (right). This method selects (53.2~$\pm$~2.3)$\%$ of events as affected by photon feedback.
The study of the 50\%-to-100\% rise-time distribution after applying the selection qualitatively supports the criterion (Fig.~\ref{fig:PF_selection}).

\section{LaB$_6$ photocathode study}
We perform an explorative study on the LaB$_6$ photocathode as a possible mitigation against ion feedback. This material is known to be more resistant to damage from both air exposure and ion feedback than CsI. It showed poor quantum efficiency (QE) at 375~nm wavelength in the 2022 laser test, but we expect better QE at Cherenkov UV energies. In addition, some changes in the photocathode configuration are applied to improve UV transmittance.
We test the GasPM performance with cosmic rays. We acquire data with both GasPM and RPC to subtract the ionisation contribution. 
The observed GasPM hit-rate of (7.19~$\pm$~0.49)\% is consistent with the (7.66~$\pm$~0.18)\% RPC rate, showing evidence for ionization-only signals. This may indicate that the quantum efficiency of the employed photocathode is too low. A LED test is planned to assess the pure photon-detection rate.

\section{Summary and prospects}

We keep improving the GasPM, a low-cost, large-coverage, $\mathcal{O}(10)$~ps-resolution RPC-based gaseous photodetector for mitigating beam background in Belle II. We identify photon and ion feedback as possible limitations. A beam test with a high-sampling-rate digitiser enables an algorithm to suppress photon-feedback. Validation of the algorithm is ongoing along with updated determination of time resolution. We also tested a more damage-resistant LaB$_6$ photocathode, which shows poor QE at the target wavelength.

\section{Acknowledgement}
This work was supported by DAIKO FOUNDATION and MEXT/JSPS KAKENHI Grant Numbers
JP26610068, JP16H00865, JP19H05099, JP21H01091, and JP23H05433. We acknowledge the support of KEK in our test at the PF-AR test beamline. S.G. acknowledges the IMAPP program and the SOKENDAI KEK Tsukuba/J-PARC Summer Student Program for the opportunity and funding.





\end{document}